\documentclass[letterpaper,spanish]{revtex4}
\usepackage[T1]{fontenc}
\usepackage[latin1]{inputenc}
\usepackage{amsmath}
\usepackage{amssymb}

\makeatletter
\usepackage{graphicx}
\usepackage{amsfonts}
\usepackage{babel}
\deactivatetilden
\makeatother
\begin{document}

\title{Agujeros de Gusano en Gravedad (2+1)}

\author{Eduard Alexis Larrañaga Rubio}

\address{Universidad Nacional de Colombia}

\address{Observatorio Astronómico Nacional (OAN)}

\begin{abstract}
Los agujeros de gusano atravesables son objetos que presentan un gran
interés en la actualidad debido a sus características geométricas
y a su relación con la materia exótica. En el presente trabajo se
muestra una revisión de las características de los agujeros de gusano
atravesables al estilo de Morris y Thorne, al igual que el proceso
de construcción y aspectos de la materia exótica necesaria para mantenerlos.
Luego, se utiliza un proceso de juntura para construir dos soluciones
específicas tipo agujero de gusano en el formalismo de la gravedad
(2+1) con constante cosmológica negativa. Con esta construcción, se
obtienen agujeros atravesables y que se encuentran unidos a un espacio-tiempo
externo correspondiente al agujero negro BTZ sin momento angular y
sin carga eléctrica. Además de esto, se muestra que para mantener
este tipo de solución es necesaria la existencia de materia exótica,
es decir, materia que viole las condiciones de energía. 
\end{abstract}
\maketitle

\section{Introducción}

Hace ya casi 20 años que aparecen los agujeros de gusano tipo Morris-Thorne\cite{morris}
como una herramienta de enseñanza de la Relatividad General. En 1995,
el conocido libro de Visser resume todos los trabajos desarrollados
en el área durante el siglo pasado. 

En la actualidad este tipo de solución de las ecuaciones de campo
de Einstein es una de las más estudiadas, debido a sus interesantes
características y en particular, debido a su relación con fuentes
de materia que violarían las condiciones de energía. Como es conocido,
desde el punto de vista topológico, los agujeros de gusano son iguales
a los agujeros negros, pero poseen una superficie minima denominada
la garganta del agujero, la cual es mantenida por una fuente de materia
que viola las condiciones de energía y por ello recibe la denominacion
\emph{exótica.} \\

Por otro lado, la gravedad $2+1$ es una teoría covariante de la geometría
del espacio-tiempo, por lo que tiene la misma base conceptual de la
Relatividad General (gravedad $3+1$), pero el modelo de dimensión
$2+1$ es mucho más simple, tanto matemática como físicamente, y por
ello se cree que puede dar algunas pistas sobre algunos aspectos cuánticos
que aún no son comprendidos. Debido a ello a recibido gran atención
en los últimos años, y aún más después la aparición de soluciones
tipo agujero negro, encontrados por Banados-Teitelboim-Zanelli (BTZ)\cite{BTZ1,BTZ2}.
Aún cuando estos son modelos de ,,juguete'' en algunos aspectos,
los agujeros BTZ siempre han tenido gran importancia debido a su conexión
con algunas teorías de cuerdas, su papel en el cómputo microscópico
de la entropía y especialmente debido a su utilidad en el estudio
de la termodinámica con correcciones cuánticas.\\
Desde hace algún tiempo se han empezado a considerar soluciones tipo
agujeor de gusano en la gravedad de $2+1$ dimensiones. Por ejemplo
Delgaty et.al. \cite{delgaty} realizan un análisis de las propiedades
de este tipo de soluciones en universos con constante cosmológica,
pero solamente considera un ejemplo específico. Por otro lado, en
el trabajo de Aminneborg et.al. \cite{Amin} se resumen muchas de
las propiedades geométricas que tienen estos objetos y son comparadas
con las propiedades de los agujeros negros. Por último, Kim et.al.
\cite{Kim} encuentran dos soluciones específicas tomando la gravedad
de dimensiñon $2+1$ en presencia de un campo dilatónico.\\

ahora bien, en este trabajo describimos los aspectos básicos de los
agujeros de gusano y de la materia exótica y se encuentran dos soluciones
en dimensionalidad $\left(2+1\right)$ siguiendo el trabajo de Lemos
et. al. \cite{Lemos}. Así, en la segunda parte se muestran las características
fundamentales de los agujeros, mientras que en la tercera se encuentran
las ecuaciones de estructura que estos deben satisfacer. En la cuarta
sección se describe el tensor momento-energía y como este ayuda a
la creación del agujero, con lo cual la quinta sección describe el
proceso de solución de las ecuaciones de campo. En la sexta parte
se describe la geometría que se espera para el agujero de gusano,
y la septima sección muestra las propiedades de la materia exótica
necesarias para su construcción, la cual se realiza explícitamente
en la octava parte. La novena parte encuentra de forma explícita dos
tipos de solución específicas y muestra la distribución de materia
necesaria para mantenerlas. En la décima sección se realiza una discusión
de los resultados obtenidos y se consignan las conclusiones.

\section{Soluciones tipo Agujero de Gusano}

\subsection{Propiedades de los Agujeros de Gusano atravesables}

Para restringir las posibles soluciones que obtendremos a los casos
de interés, debemos imponer ciertas propiedades que deben poseer las
soluciones tipo agujero de gusano\cite{morris}. Estas propiedades
son:

\begin{enumerate}
\item La métrica debe ser solución de las ecuaciones de campo  en todo punto
del espacio-tiempo.
\item La métrica debe ser esféricamente simétrica y debe ser estática (i.e.
debe ser independiente del tiempo.
\item La solución debe poseer una ''garganta'' que conecte dos regiones
del espacio-tiempo que para el caso de la gravedad $\left(2+1\right)$
deben corresponder a la métrica del agujero BTZ.
\item En la métrica no debe existir un horizonte de sucesos, ya que este
impediría el viaje en dos direcciones dentro del agujero.
\item La fuerza de marea gravitacional dentro del agujero debe ser pequeña
(para permitir el viaje de un observador).
\item El tiempo necesario para cruzar el agujero de gusano debe ser razonable. 
\end{enumerate}
Estas condiciones se impondran a lo largo del desarrollo para obtener
finalmente las soluciones específicas.

\section{Ecuaciones de estructura del Agujero de Gusano}

\bigskip

\subsection{Ecuaciones de Campo de la gravedad 2+1 con Constante Cosmológica}

Debemos asegurar que la métrica que describirá el agujero de gusano
es solución de las ecuaciones de campo en todo punto del espacio-tiempo.
Ya que deseamos obtener soluciones con constante cosmológica, las
ecuaciones de campo estarán dadas por:

\bigskip\begin{equation}
G_{\mu v}+\Lambda g_{\mu v}=T_{\mu v}\label{eccampo}\end{equation}

donde $g_{\mu v}$ es la métrica, $\Lambda$ es la constante cosmológica,
$T_{\mu v}$ es el tensor momento-energía y $G_{\mu v}$ es el tensor
de Einstein, definido por:\begin{equation}
G_{\mu v}=R_{\mu v}-\frac{1}{2}g_{\mu v}R\end{equation}

con $R_{\mu v}$ el tensor de Ricci, construido como la contracción
del tensor de Riemann (curvatura): $R_{\mu v}=R_{\mu\alpha v}^{\alpha}$;
y $R$ es el escalar de curvatura, construido a partir de la contracción
del tensor de Ricci: $R=R_{\alpha}^{\alpha}$. Los índices griegos
toman los valores $0,1,2$.

\bigskip

\subsection{La Métrica}

La métrica que va representar el agujero de gusano debe poseer una
simetría esférica, y además debe ser independiente del tiempo. Esto
hace que la forma del elemento de línea, utilizando las coordenadas
esféricas usuales $\left(t,r,\varphi\right)$, sea\begin{equation}
ds^{2}=-e^{2\Phi\left(r\right)}dt^{2}+\frac{1}{1-\frac{b\left(r\right)}{r}}dr^{2}+r^{2}d\varphi^{2},\label{metrica}\end{equation}

donde $\Phi\left(r\right)$ , $b\left(r\right)$ son funciones arbitrarias
que dependen únicamente de la coordenada radial $r$, y que estarán
restringidas por las condiciones que impondremos a la solución. Como
se verá más adelante, la función $\Phi\left(r\right)$ determinará
el corrimiento al rojo gravitacional, por lo que en la literatura
se conoce como ''\textit{función de} \emph{redshift}''. Por otro
lado, la función $b\left(r\right)$ determina la forma espacial de
agujero de gusano, por lo que recibe el nombre de ''\textit{función
de forma}''

\subsection{Tensor de Einstein para la métrica del agujero de gusano}

El primer paso para llegar a las ecuaciones de campo es el determinar
el tensor de Riemann, el tensor de Ricci y el escalar de curvatura
a partir de la métrica dada por la ecuación (\ref{metrica}). En el
Apéndice A se encuentran explícitamente las componentes no nulas de
estos tensores. Finalmente, se obtienen 3 componentes diferentes de
cero para el Tensor de Einstein,

\bigskip\begin{align}
G_{tt} & =\frac{1}{2r^{3}}e^{2\Phi}\left[-b+rb^{\prime}\right]\label{einstein1}\\
G_{rr} & =\frac{\Phi^{\prime}}{r}\nonumber \\
G_{\varphi\varphi} & =\frac{1}{2}\left[\Phi^{\prime}\left(b-rb^{\prime}\right)+2r\left(r-b\right)\left(\left(\Phi^{\prime}\right)^{2}+\Phi^{\prime\prime}\right)\right],\nonumber \end{align}

donde hemos denotado con primas las derivadas con respecto a la coordenada
radial $r$.

\bigskip

\subsection{Cambio de Base}

Debemos notar que el tensor de Einstein calculado en el parágrafo
anterior está en una base vectorial (tetrada) dada por $\left(\mathbf{e}_{t},\mathbf{e}_{r},\mathbf{e}_{\varphi}\right)$,
que esta asociada con el sistema de coordenadas $t,r,\varphi$. Sin
embargo, sabemos que somos libres de elegir el sistema de referencia
que nosotros deseemos.

De esta manera, para los cálculos subsecuentes, al igual que para
las interpretaciones físicas, es más conveniente utilizar como base
un conjunto de vectores ortonormales. Estos corresponderán al sistema
de referencia propio de un conjunto de observadores que permanezcan
en reposo en el sistema coordenado $\left(t,r,\varphi\right)$; es
decir con $r,\varphi$ constantes.

Denotaremos los vectores de la base ortonormal por $\left(\mathbf{e}_{\widehat{t}},\mathbf{e}_{\widehat{r}},\mathbf{e}_{\widehat{\varphi}}\right)$;
y estarán expresados en términos de los vectores de la base coordenada
$\left(\mathbf{e}_{t},\mathbf{e}_{r},\mathbf{e}_{\varphi}\right)$
mediante la transformación\begin{equation}
\mathbf{e}_{\widehat{\alpha}}=\Lambda_{\widehat{\alpha}}^{\beta}\mathbf{e}_{\beta}\end{equation}

donde la matriz de transformación está dada por\begin{equation}
\Lambda_{\widehat{\alpha}}^{\beta}=diag\left[e^{-\Phi},\left(1-\frac{b}{r}\right)^{1/2},\frac{1}{r}\right].\end{equation}

Explícitamente esta transformación será\begin{align}
\mathbf{e}_{\widehat{t}} & =e^{-\Phi}\mathbf{e}_{t}\\
\mathbf{e}_{\widehat{r}} & =\left(1-\frac{b}{r}\right)^{1/2}\mathbf{e}_{r}\\
\mathbf{e}_{\widehat{\varphi}} & =\frac{1}{r}\mathbf{e}_{\varphi}\end{align}

Es importante notar que en esta base, el tensor métrico tomará la
forma de la relatividad especial, es decir, corresponde al tensor
de Minkowski:\begin{equation}
g_{\alpha\beta}=\mathbf{e}_{\widehat{\alpha}}\cdot\mathbf{e}_{\widehat{\beta}}=\eta_{\widehat{\alpha}\widehat{\beta}}=\left[\begin{array}{ccc}
-1 & 0 & 0\\
0 & 1 & 0\\
0 & 0 & 1\end{array}\right]\end{equation}

Al cambiar de sistema de referencia, las componentes de todos nuestros
tensores cambiarán. De esta manera, por ejemplo, en el sistema de
referencia ortonormal las ecuaciones de Einstein con constante cosmológica
estarán dadas por\begin{equation}
G_{\widehat{\mu}\widehat{v}}+\Lambda\eta_{\widehat{\mu}\widehat{v}}=T_{\widehat{\mu}\widehat{v}}.\label{eccampo2}\end{equation}

\bigskip

El tensor de Einstein dado por las ecuaciones (\ref{einstein1}) tendrá
en la base ortonormal la forma más simple (ver Apéndice B para cálculos
explícitos):\begin{align}
G_{\widehat{t}\widehat{t}} & =\frac{1}{2r^{3}}\left[b^{\prime}r-b\right]\label{teinstein1}\\
G_{\widehat{r}\widehat{r}} & =\left(1-\frac{b}{r}\right)\frac{\Phi^{\prime}}{r}\\
G_{\widehat{\varphi}\widehat{\varphi}} & =\frac{1}{2r^{2}}\left[\Phi^{\prime}\left(b-rb^{\prime}\right)+2r\left(r-b\right)\left(\left(\Phi^{\prime}\right)^{2}+\Phi^{\prime\prime}\right)\right]\label{teinstein3}\end{align}

\bigskip

\section{El Tensor Momento-Energía}

Para construir un agujero de gusano con las propiedades de traversabilidad
que queremos, debe existir un tensor momento energía diferente de
cero.

Las ecuaciones de campo (\ref{eccampo2}) nos dicen que el tensor
momento energía es proporcional al tensor de Einstein. Es decir que
el tensor $T_{\widehat{\mu}\widehat{v}}$ debe tener la misma estructura
algebraica que $G_{\widehat{\mu}\widehat{v}}$; por lo que las únicas
componentes que son diferentes de cero deben ser: $T_{\widehat{t}\widehat{t}},T_{\widehat{r}\widehat{r}}$
y $T_{\widehat{\varphi}\widehat{\varphi}}$.

Ahora bien, ya que estamos trabajando en una base ortonormal $\left(\mathbf{e}_{\widehat{t}},\mathbf{e}_{\widehat{r}},\mathbf{e}_{\widehat{\varphi}}\right)$
relacionada con el sistema de referencia inercial para los observadores
estáticos, estas componentes del tensor momento-energía tienen una
interpretación inmediata\begin{align}
T_{\widehat{t}\widehat{t}} & =\rho\left(r\right)\label{tmomento}\\
T_{\widehat{r}\widehat{r}} & =-\tau\left(r\right)\nonumber \\
T_{\widehat{\varphi}\widehat{\varphi}} & =p\left(r\right),\nonumber \end{align}

donde $\rho\left(r\right)$ es la densidad total de masa y energía
(en unidades de $\operatorname{kg}/\operatorname{m}^{3}$); $\tau\left(r\right)$
es la tensión radial por unidad de área, es decir es el negativo de
la presión radial, $\tau\left(r\right)=-p_{r}\left(r\right)$ (en
unidades de $\operatorname{N}/\operatorname{m}^{2}$); y $p\left(r\right)$
es la presión medida en las direcciones tangenciales, i.e. ortogonales
a la dirección radial (en unidades de $\operatorname{N}/\operatorname{m}^{2}$).

\subsection{La Constante Cosmológica como parte del Tensor Momento-Energía}

Las ecuaciones de campo de Einstein con constante cosmol\'{o}gica
(\ref{eccampo2}) son\begin{equation}
G_{\widehat{\mu}\widehat{v}}+\Lambda\eta_{\widehat{\mu}\widehat{v}}=T_{\widehat{\mu}\widehat{v}}.\end{equation}

Para obtener una interpretación del término cosmológico podemos definir
un tensor $T_{\widehat{\mu}\widehat{v}}^{\left(vac\right)}$ como\begin{equation}
T_{\widehat{\mu}\widehat{v}}^{\left(vac\right)}=-\Lambda\eta_{\widehat{\mu}\widehat{v}}=\left[\begin{array}{ccc}
\Lambda & 0 & 0\\
0 & -\Lambda & 0\\
0 & 0 & -\Lambda\end{array}\right],\end{equation}

con lo cual podemos reescribir las ecuaciones de campo como\begin{equation}
G_{\widehat{\mu}\widehat{v}}=\left(T_{\widehat{\mu}\widehat{v}}+T_{\widehat{\mu}\widehat{v}}^{\left(vac\right)}\right)\end{equation}
\begin{equation}
G_{\widehat{\mu}\widehat{v}}=\overline{T}_{\widehat{\mu}\widehat{v}},\end{equation}

donde $\,\,\overline{T}_{\widehat{\mu}\widehat{v}}=T_{\widehat{\mu}\widehat{v}}+T_{\widehat{\mu}\widehat{v}}^{\left(vac\right)}$
es el tensor momento-energía total. Es evidente ahora que podemos
interpretar $T_{\widehat{\mu}\widehat{v}}^{\left(vac\right)}$ como
el tensor momento-energ\'{\i}a asociado con el vacio.

De esta manera, las funciones $\overline{\rho}\left(r\right),\overline{\tau}\left(r\right)$
y $\overline{p}\left(r\right)$ totales estarán dadas por:\begin{align}
\overline{\rho}\left(r\right) & =\rho\left(r\right)+\Lambda\\
\overline{\tau}\left(r\right) & =\tau\left(r\right)+\Lambda\nonumber \\
\overline{p}\left(r\right) & =p\left(r\right)-\Lambda\nonumber \end{align}

\section{Solución de las Ecuaciones de Campo}

Utilizando la forma explícita del tensor de Einstein dado por (\ref{teinstein1}-\ref{teinstein3})
y del tensor momento-energía (\ref{tmomento}), podemos reemplazar
en las ecuaciones de campo de Einstein, para obtener las ecuaciones:\begin{equation}
\rho\left(r\right)=\frac{1}{2r^{3}}\left[b^{\prime}r-b\right]-\Lambda\label{aux1}\end{equation}
\begin{equation}
\tau\left(r\right)=-\left(1-\frac{b}{r}\right)\frac{\Phi^{\prime}}{r}-\Lambda\label{aux2}\end{equation}
\begin{equation}
p\left(r\right)=\frac{1}{2r^{2}}\left[\Phi^{\prime}\left(b-rb^{\prime}\right)+2r\left(r-b\right)\left(\left(\Phi^{\prime}\right)^{2}+\Phi^{\prime\prime}\right)\right]+\Lambda\label{aux3}\end{equation}

Si tomamos la derivada con respecto a la coordenada $r$ en la ecuaci\'{o}n
\ref{aux2} tendremos:\begin{align*}
\tau^{\prime}\left(r\right) & =-\left(1-\frac{b}{r}\right)\frac{\Phi^{\prime\prime}}{r}+\left(1-\frac{b}{r}\right)\frac{\Phi^{\prime}}{r^{2}}+\frac{b^{\prime}r-b}{r^{3}}\Phi^{\prime}\end{align*}

Utilizando las ecuaciones \ref{aux1}-\ref{aux3} para eliminar $b^{\prime}$
y $\Phi^{\prime\prime}$ obtenemos

\begin{align}
\tau^{\prime}\left(r\right) & =-\left(1-\frac{b}{r}\right)\frac{\Phi^{\prime\prime}}{r}+\left(1-\frac{b}{r}\right)\frac{\Phi^{\prime}}{r^{2}}+2\left(\rho+\Lambda\right)\Phi^{\prime}\\
\tau^{\prime}\left(r\right) & =-\frac{p-\Lambda}{r}-\left(\rho+\Lambda\right)\Phi^{\prime}+\left(1-\frac{b}{r}\right)\frac{\left(\Phi^{\prime}\right)^{2}}{r}+\left(1-\frac{b}{r}\right)\frac{\Phi^{\prime}}{r^{2}}+2\left(\rho+\Lambda\right)\Phi^{\prime}\\
\tau^{\prime}\left(r\right) & =-\frac{p+\left(1-\frac{b}{r}\right)\frac{\Phi^{\prime}}{r}+\tau}{r}+\left(1-\frac{b}{r}\right)\frac{\left(\Phi^{\prime}\right)^{2}}{r}+\left(1-\frac{b}{r}\right)\frac{\Phi^{\prime}}{r^{2}}+\left(\rho-\left(1-\frac{b}{r}\right)\frac{\Phi^{\prime}}{r}-\tau\right)\Phi^{\prime}\end{align}
\begin{equation}
\tau^{\prime}\left(r\right)=\left(\rho-\tau\right)\Phi^{\prime}-\frac{p+\tau}{r}\label{aux4}\end{equation}

Las ecuaciones (\ref{aux1}), (\ref{aux3}) y (\ref{aux4}) son tres
ecuaciones diferenciales que relacionan las cinco funciones desconocidas:
$b,\Phi,\rho,\tau$ y $p$.

Ahora bien, la forma usual de resolver estas ecuaciones es asumir
una tipo específico de materia y energía. Con la ecuación de estado
correspondiente al tipo de material se encuentran la tensión en funci\'{o}n
de la densidad $\tau\left(\rho\right)$ y la presión en función de
la densidad $p\left(\rho\right)$. Así, las dos ecuaciones de estado
mas las tres ecuaciones de campo serán cinco ecuaciones para cinco
incógnitas. De esta manera, se encontraría la forma que tendr\'{a}
el espacio-tiempo, es decir las funciones $b\left(r\right)$ y $\Phi\left(r\right)$.

\bigskip

En el caso de los agujeros de gusano procederemos de una manera diferente.
Ya que hemos impuesto ciertas condiciones que deseamos que tenga nuestro
espacio-tiempo parta describir un agujero de gusano, entonces controlaremos
las funciones $b\left(r\right)$ y $\Phi\left(r\right)$, fijandolas
para una geometría conveniente, y, utilizando las ecuaciones de campo,
encontraremos la distribución de materia-energía necesaria para obtener
tal solución.

\bigskip

\section{Geometría del Agujero de Gusano}

Antes de continuar con la construcción específica de soluciones tipo
agujeros de gusano debemos introducir algunos elementos que nos permitan
analizar las cualidades de la geometría que deseamos.

\subsection{La Matemática de las Inmersiones}

Ahora utilizaremos los diagramas de inmersión para representar el
agujero de gusano y con ello poder obtener información acerca de cómo
debe ser la función de forma $b\left(r\right)$. Consideremos la métrica
del agujero de gusano, dada por la ecuación (\ref{metrica}), para
un tiempo fijo $t$,\begin{equation}
ds^{2}=\frac{dr^{2}}{1-\frac{b}{r}}+r^{2}d\varphi^{2}.\label{corte}\end{equation}

Lo que vamos a hacer ahora es construir, en el espacio euclideano
tridimensional, una superficie bidimensional con la misma geometría
de este corte; es decir que realizaremos una \textit{inmersión} del
corte dentro del espacio euclideano.

Para ello consideremos la métrica euclideana en coordenadas cilíndricas\begin{equation}
ds^{2}=dz^{2}+dr^{2}+r^{2}d\varphi^{2}\end{equation}
 \bigskip

La superficie bidimensional tendrá simetría cilíndrica, y por lo tanto
podrá describirse por una función $z=z\left(r\right)$. Por lo tanto
la métrica de esa superficie se puede escribir como:\begin{equation}
ds^{2}=\left[1+\left(\frac{dz}{dr}\right)²\right]dr^{2}+r^{2}d\varphi^{2}\end{equation}

y este elemento de línea debe ser el mismo que el del corte ecuatorial
dado por (\ref{corte}). De esta manera, para realizar la inmersión,
la función $z\left(r\right)$ debe satisfacer:\begin{equation}
1+\left(\frac{dz}{dr}\right)²=\left(1-\frac{b}{r}\right)^{-1}\end{equation}
\begin{equation}
\frac{dz}{dr}=\pm\left(\frac{r}{b}-1\right)^{-1/2}\label{aux5}\end{equation}

Para que esta geometría describa un agujero de gusano, la solución
debe poseer un radio mínimo $r=b\left(r\right)=r_{m}$, que llamaremos
la ''garganta'' del agujero, y para el cual la superficie inmersa
es vertical, es decir que en este punto $\frac{dz}{dr}\rightarrow\infty$.
Debido a esta divergencia de $\frac{dz}{dr}$, la coordenada $r$
no es la apropiada cerca a la garganta. Por ello es mejor utilizar
la distancia radial propia (medida por observadores estáticos), definida
por\begin{equation}
l\left(r\right)=\pm\int_{r_{m}}^{r}\frac{dr}{\sqrt{1-\frac{b}{r}}}\label{aux6}\end{equation}

donde el signo $+$ se utiliza para la parte superior del agujero
(i.e. $z>0$) y el signo $-$ para la parte inferior ($z<0$). El
límite superior máximo en la integración corresponderá al radio $a$
en el cuál se situa la boca del agujero de gusano.

Ya que la distancia radial propia $l\left(r\right)$ debe permanecer
finita en todo punto, debemos exigir que\begin{equation}
1-\frac{b\left(r\right)}{r}\geq0.\end{equation}

Las ecuaciones (\ref{aux5}) y (\ref{aux6}) implican que para la
superficie inmersa se cumple\begin{align}
\frac{dz}{dl} & =\pm\sqrt{\frac{b}{r}}\\
\frac{dr}{dl} & =\pm\sqrt{1-\frac{b}{r}}.\end{align}

De esta manera, la superficie $z\left(r\right)$ puede tener una forma
general, y la función $b\left(r\right)$, en efecto, define la forma
del agujero de gusano.

\bigskip

\subsection{Ausencia de un Horizonte}

Para asegurar que el agujero de gusano permita el viaje hacia adentro
y hacia afuera, debemos exigir que no existan horizontes de eventos.
En el caso de métricas estáticas, es fácil reconocer la presencia
de tales horizontes: estos corresponderán a las superficies no singulares
en las cuales\[
g_{tt}=-e^{2\Phi}\rightarrow0,\]

es decir, equellos puntos donde el intervalo de tiempo propio es nulo
mientras transcurre un tiempo coordenado finito. De esta manera, para
asegurar el viaje en las dos direcciones debemos exigir que la función
$\Phi\left(r\right)$ \textit{sea finita en todo punto.}

\section[Tensor Momento-Energía]{Propiedades del Tensor Momento-Energía que genera el Agujero de Gusano}

Como hemos visto, el método que se utiliza para solucionar las ecuaciones
de campo en el contexto de los agujeros de gusano es diferente al
usual. En este caso impondremos ciertas condiciones a la función de
forma $b\left(r\right)$, y utilizando las ecuaciones de campo (\ref{aux1})
a (\ref{aux3}) se obtienen ciertas condiciones sobre la densidad
$\rho$, la tensión radial $\tau$ y la presión lateral $p$, de tal
manera que se genera el espacio-tiempo deseado. Sin embargo, esta
forma de actuar nos llevará a la necesidad de utilizar un tipo muy
particular de materia, el cuál denominaremos ''exótica'' debido
a sus propiedades.

\bigskip

\subsection{Materia Exótica}

Para comprobar qué propiedades debe cumplir el tensor momento-energía
necesario para lograr la configuración de agujero de gusano, definiremos
la siguiente función adimensional\begin{equation}
\varsigma=\frac{\tau-\rho}{\left|\rho\right|}.\end{equation}

Utilizando las ecuaciones (\ref{aux1}) y (\ref{aux2}) esta función
puede escribirse como:\begin{equation}
\varsigma=\frac{\tau-\rho}{\left|\rho\right|}=\frac{-2r^{2}\left(1-\frac{b}{r}\right)\Phi^{\prime}-\left(b^{\prime}r-b\right)}{\left|b^{\prime}r-b-2r^{3}\Lambda\right|}\label{aux22}\end{equation}

Para que el espacio-tiempo descrito corresponda a un agujero de gusano
debe exigirse que este se conecte suavemente con el espacio-tiempo
exterior (el cual en nuestro caso es asintóticamente AdS ). Esto significa
que debemos exigir que la garganta de la superficie inmersa debe ''abrirse''
hacia afuera, tal como se observa en las gráficas que representan
agujeros de gusano.

Ahora bien, esta condición de ''abrirse'' hacia afuera se representa
matemáticamente imponiendo al inverso de la función de inmersión $r\left(z\right)$
la condición\begin{equation}
\frac{d^{2}r}{dz^{2}}>0.\end{equation}

\bigskip Utilizando la ecuación (\ref{aux5}) tenemos\begin{equation}
\frac{dr}{dz}=\pm\left(\frac{r}{b}-1\right)^{1/2},\end{equation}

y diferenciando con respecto a $z$ obtenemos\begin{equation}
\frac{d^{2}r}{dz^{2}}=\pm\frac{1}{2}\left(\frac{b-rb^{\prime}}{b^{2}}\right).\end{equation}

Es decir que la condición de ''abrirse'' hacia afuera se puede escribir
como\begin{equation}
\frac{d^{2}r}{dz^{2}}=\frac{b-rb^{\prime}}{2b^{2}}>0\text{ \,\,\,\, en la garganta o cerca de ella.}\end{equation}

\bigskip De esta manera, la ecuación (\ref{aux22}) será\begin{equation}
\varsigma=\frac{\tau-\rho}{\left|\rho\right|}=\frac{2b^{2}}{\left|2b^{2}\frac{d^{2}r}{dz^{2}}+2r^{3}\Lambda\right|}\frac{d^{2}r}{dz^{2}}-2\left(1-\frac{b}{r}\right)\frac{r^{2}\Phi^{\prime}}{\left|2b^{2}\frac{d^{2}r}{dz^{2}}+2r^{3}\Lambda\right|}.\end{equation}

Ahora bien, sabemos que cerca a la garganta tenemos $\left(1-\frac{b}{r}\right)\Phi^{\prime}\longrightarrow0$,
y por ello, la condición de ''abrirse'' hacia afuera se escribir\'{a}
ahora como\begin{equation}
\varsigma_{m}=\frac{\tau_{m}-\rho_{m}}{\left|\rho_{m}\right|}>0,\label{aux23}\end{equation}

donde el subindice $_{m}$ muestra que estamos evaluando la expresión
en la garganta o cerca de ella.

La condición $\tau_{m}>\rho_{m}$ que impone (\ref{aux23}) nos dice
que, en la garganta, la tensión $\tau_{m}$ debe ser mayor que la
densidad total de masa-energía $\rho_{m}$. Cualquier material que
cumpla esta propiedad $\left(\tau_{m}>\rho_{m}>0\right)$ será llamado
''exótico''. Para comprender por qué esta propiedad es problemática,
debemos repasar las llamadas \textit{condiciones de energía}.

\bigskip

\subsection{Condiciones de Energía}

Durante los años 60's y 70's una de las conjeturas más fuertemente
arraigadas dentro de la física era la de que ningún observador puede
medir una densidad de energía negativa. Esta conjetura es la que actualmente
se conoce como \textit{condición de energía débil} (WEC). Matemáticamente
podemos escribir esta condición diciendo que el tensor momento-energía
siempre satisface\cite{hawking}\begin{equation}
T_{\mu v}W^{\mu}W^{v}\geq0,\label{WEC}\end{equation}

para cualquier vector como de tiempo $W^{\mu}$.

A la par con esta condición, surge la denominada \textit{condición
de energía dominante} (DEC), la cual puede interpretarse diciendo
que ningún observador puede medir energ\'{\i}as negativas y además
el vector de flujo de energía local nunca es como de espacio. Matemáticamente
esta condición dice que para cualquier vector como de tiempo $W^{\mu}$
se cumple\begin{equation}
T_{\mu v}W^{\mu}W^{v}\geq0\text{ \,\,\,\,\,\,\,\,\,\, y \,\,\,\,\,\,}T^{\mu v}W_{\mu}\text{ \,\, NO es un vector como de espacio. }\label{DEC}\end{equation}

Por último, también se tiene la llamada \textit{condición de energía
fuerte} (SEC) que impone la condición\begin{equation}
T_{\mu v}W^{\mu}W^{v}\geq\frac{1}{2}TW^{\sigma}W_{\sigma}\label{SEC}\end{equation}

para cualquier vector como de tiempo $W^{\mu}$ y donde $T$ es la
traza de $T_{\mu v}$.

\bigskip

Existe además otra condición, denominada \textit{condición de energía
nula} (NEC), introducida un poco después y que se escribe matemáticamente
diciendo que para cualquier vector nulo $V^{\mu}$ se cumple\begin{equation}
T_{\mu v}V^{\mu}V^{v}\geq0.\label{NEC}\end{equation}

\bigskip

Para visualizar mejor estas condiciones de energía tomemos el caso
particular de un fluido perfecto con tensor de momento energía diagonal
$T_{\mu v}=diag\left(\rho,-p-p-p\right)$. En este caso, las condiciones
de energía serán (ver pág. 81 de \cite{bangs})\begin{align}
NEC & \Longrightarrow\left(\rho+p\right)\geq0\\
WEC & \Longrightarrow\rho\geq0\text{ \,\,\,\, y \,\,\,\,}\left(\rho+p\right)\geq0\\
SEC & \Longrightarrow\left(\rho+3p\right)\geq0\text{ \,\, y \,\,}\left(\rho+p\right)\geq0\\
DEC & \Longrightarrow\rho\geq0\text{ \,\, y \,\,}\left(\rho\pm p\right)\geq0.\end{align}

De estas ecuaciones se puede observar claramente que las condiciones
de energía son en realidad relaciones lineales entre la densidad $\rho$
y la presión $p$ de la materia-energía que genera la curvatura del
espacio-tiempo. Además, es importante notar que estas relaciones no
son demostrables, ya que son solo conjeturas que suenan ''razonables''
físicamente\cite{hawking} debido a que la masa y la mayoría de los
campos que se conocen las cumplen.

\bigskip

\subsection{Agujeros de Gusano y las Condiciones de Energía}

\bigskip

La naturaleza exótica del material que produce un agujero de gusano,
dada por la condición $\tau_{m}>\rho_{m},$ es problemática debido
a que conlleva a la violación de las condiciones de energía.

Consideremos un observador que se mueve a traves de la garganta de
un agujero de gusano con una velocidad $v$ muy grande $\left(\gamma>>1\right)$.
La densidad de energía que él observa es la proyección del tensor
momento-energía (dado por \ref{tmomento}) con su vector de base temporal
$\mathbf{e}_{\widehat{o}}=\gamma\mathbf{e}_{\widehat{t}}\mp\gamma\left(\frac{v}{c}\right)\mathbf{e}_{\widehat{r}}$.
De esta manera, la densidad estará dada por:\begin{align}
T_{\widehat{o}\widehat{o}} & =\gamma^{2}T_{\widehat{t}\widehat{t}}\mp2\gamma^{2}\left(\frac{{\normalsize v}}{c}\right)^{2}T_{\widehat{t}\widehat{r}}+\gamma^{2}\left(\frac{{\normalsize v}}{c}\right)^{2}T_{\widehat{r}\widehat{r}}\\
 & =\gamma^{2}\left[\rho_{m}c^{2}-\left(\frac{{\normalsize v}}{c}\right)^{2}\tau_{m}\right]\\
 & =\gamma^{2}\left(\rho_{m}c^{2}-\tau_{m}\right)+\tau_{m}\end{align}

Por lo tanto, si se cumple $\tau_{m}>\rho_{m}$, y el observador se
mueve lo suficientemente rápido (i.e. con un $\gamma$ suficientemente
grand), entonces éste podrá medir una densidad de masa-energía negativa.

\subsection{Violación de las Condiciones de Energía}

\bigskip

Como se dijo antes, las condiciones de enería no son demostrables,
son solamente conjeturas. Sin embargo, son fundamentales en la demostración
de ciertos teoremas importantes como por ejemplo en el teorema de
''masa positiva'', el cual dice que cualquier objeto compuesto de
materia que cumpla la DEC, nunca antigravita. También son utilizadas
en ciertos teoremas que aseguran la creación de singularidades en
situaciones cosmológicas \cite{hawking} y en el teorema del ''Censor
C\'{o}smico''\cite{bangs}.

\bigskip

Aún más importante es que las condiciones de energía también son utilizadas
en la demostración del teorema del ''aumento de área'' conocido
como la ''segunda ley de la termodinámica'' de agujeros negros.
Este nos asegura que si toda la materia-energía alrededor de un agujero
negro satisface la SEC, entonces el área de horizonte de eventos del
agujero nunca decrece\cite{hawking2}.

Sin embargo, el mismo Hawking demostró que los agujeros negros no-rotantes
pueden evaporarse, y por lo tanto su área decrecerá, violando así
la segunda ley\cite{hawking3,hawking4}. Este hecho hace evidente
la posibilidad de que los campos cuánticos violen las condiciones
de energía. Más exactamente, lo que se obtiene es que pueden existir
ciertos estados cuánticos en los que el valor esperado renormalizado
del tensor momento-energía viola las condiciones de energía.\\

El ejemplo más claro de este hecho es el proceso de creación de partículas.
Por ejemplo, cualquier observador estático justo afuera del horizonte
de eventos de un agujero de Schwarzschild (rodeado por vacio) ve un
valor esperado para la densidad de energía negativo. Esta densidad
negativa de energía es la que se asocia precisamente con la creación
de partículas cerca al horizonte, las cuales a su vez son las responsables
del flujo de energía negativo hacia el horizonte, que tiene como consecuencia
la evaporación del agujero negro.

\bigskip

Otra situación bastante conocida en la cual se violan las condicions
de energía es el denominado Efecto Casimir, en el cual se considera
la energía del vacio asociada con un campo electromagnético en la
región entre dos placas reflectoras planas paralelas\cite{birrell}.
Debido a la condición de frontera impuesta por la presencia de las
placas los modos posibles para el campo electromagnético están restringidos
y el campo sufre una distorsión topológica que conlleva a un valor
esperado del tensor momento-energía negativo. Debido a que este valor
esperado no es nulo, aparece una fuerza de atracción entre las dos
placas conductoras aún cuando estas sean eléctricamente neutras. El
efecto Casimir pudo ser detectado experimentalmente por Sparnaay (1958)
y Tabor y Winterton (1969); lo que muestra de nuevo la violación de
las condiciones de energía.

Utilizando este efecto, Morris et.al.\cite{morris2} muestran como
un campo eléctrico o magnético radial cerca de la boca de un agujero
de gusano se comporta casi de manera exótica.

\bigskip

Ahora bien, aún cuando las observaciones han mostrado violaciones
producidas en pequeños sistemas cuánticos, hoy en día no es claro
si pueden existir violaciones de las condiciones de energía a nivel
macroscópico. Así, estas observaciones sirven para advertirnos que
no debemos asumir que es imposible la existencia del material ''exótico''
necesario para mantener un agujero de gusano. De hecho, la existencia
de materia con masa gravitacional negativa es una posibilidad que
se ha estudiado repetidamente en la historia de la física, y la no
observación de esta clase de materia en nuestra localidad del universo
puede ser explicada al asumir que debido al tipo de interacción que
tendría con la materia ordinaria (gravitacionalmente positiva), la
materia gravitacional negativa habría sido expulsada a distancia extragalacticas.

\section[Construcción de los Agujeros de Gusano]{Construcción de los Agujeros de Gusano con Constante Cosmológica}

Para realizar la construcción de los agujeros de gusano utilizaremos
las ecuaciones que relacionan las funciones $b,\Phi,\rho,\tau$ y
$p$,\begin{align}
\rho\left(r\right) & =\frac{1}{2r^{3}}\left[b^{\prime}r-b\right]-\Lambda\label{aux7}\\
\tau\left(r\right) & =-\left(1-\frac{b}{r}\right)\frac{\Phi^{\prime}}{r}-\Lambda\nonumber \\
p\left(r\right) & =\frac{1}{2r^{2}}\left[\Phi^{\prime}\left(b-rb^{\prime}\right)+2r\left(r-b\right)\left(\left(\Phi^{\prime}\right)^{2}+\Phi^{\prime\prime}\right)\right]+\Lambda\nonumber \\
\tau^{\prime}\left(r\right) & =\left(\rho-\tau\right)\Phi^{\prime}-\frac{p+\tau}{r}.\nonumber \end{align}

Ya que trabajaremos con el término de constante cosmológica, debemos
distinguir entre la solución interior, ( i.e. $r<a$, con $\Lambda_{int}$)
y la solución exterior (i.e. $r>a$, con $\Lambda_{ext}$).

\subsection{Solución Interior}

La solución interior debe tener la forma\begin{equation}
ds^{2}=-e^{2\Phi^{int}\left(r\right)}c^{2}dt^{2}+\frac{1}{1-\frac{b^{int}\left(r\right)}{r}}dr^{2}+r^{2}d\varphi^{2}.\label{metricainterior}\end{equation}

\bigskip

Para encontrar explícitamente las funciones $\Phi_{int}\left(r\right)$
y $b_{int}\left(r\right)$ en el interior, es decir para radios $r<a$,
en las ecuaciones (\ref{aux7}) debemos utilizar $\Lambda_{int}$.
Para no sobrecargar las ecuaciones no colocaremos el índice $^{int}$
en $\Phi$ y $b$. De esta manera tenemos:\begin{equation}
\rho\left(r\right)=\frac{1}{2r^{3}}\left[b^{\prime}r-b\right]-\Lambda_{int}\label{aux14}\end{equation}
\begin{equation}
\tau\left(r\right)=-\left(1-\frac{b}{r}\right)\frac{\Phi^{\prime}}{r}-\Lambda_{int}\label{aux13}\end{equation}
\begin{equation}
p\left(r\right)=\frac{1}{2r^{2}}\left[\Phi^{\prime}\left(b-rb^{\prime}\right)+2r\left(r-b\right)\left(\left(\Phi^{\prime}\right)^{2}+\Phi^{\prime\prime}\right)\right]+\Lambda_{int}\label{aux15}\end{equation}

De la ecuación para la tensión, vemos que en la garganta del agujero
($b\left(r_{m}\right)=r_{m}$), se tiene

\bigskip\begin{equation}
\tau\left(r_{m}\right)=-\left(1-\frac{r_{m}}{r_{m}}\right)\frac{\Phi^{\prime}}{r_{m}}-\Lambda_{int}\end{equation}
\begin{equation}
\tau\left(r_{m}\right)=-\Lambda_{int}.\end{equation}

Es decir que la tensión radial en la garganta es positiva para agujeros
en los cuales la estructura interna sea tal que $\Lambda_{int}<0$. 

De manera similar, la tensión en la garganta es negativa, es decir
es una presión, para agujeros de gusano en los cuales la estructura
interna sea tal que $\Lambda_{int}>0$.

Por otro lado, la tensi\'{o}n radial total en la garganta es nula!,
\begin{equation}
\overline{\tau}\left(r_{m}\right)=\tau\left(r_{m}\right)+\Lambda_{int}=0.\end{equation}

\subsection{Solución Exterior}

En la parte exterior del agujero de gusano ($r>a$) consideraremos
una geometría para un espacio-tiempo vacio; es decir que el tensor
momento-energía en el exterior será nulo, $T_{\widehat{\mu}\widehat{v}}=0$,
lo que implica directamente que

\begin{equation}
\rho\left(r\right)=\tau\left(r\right)=p\left(r\right)=0.\end{equation}

Sin embargo, esto no implica que la constante cosmológica exterior
$\Lambda_{ext}$ sea nula. Las ecuaciones (\ref{aux7}) serán ahora\begin{align}
0 & =\frac{1}{2r^{3}}\left[b^{\prime}r-b\right]-\Lambda_{ext}\\
0 & =-\left(1-\frac{b}{r}\right)\frac{\Phi^{\prime}}{r}-\Lambda_{ext}\nonumber \\
0 & =\frac{1}{2r^{2}}\left[\Phi^{\prime}\left(b-rb^{\prime}\right)+2r\left(r-b\right)\left(\left(\Phi^{\prime}\right)^{2}+\Phi^{\prime\prime}\right)\right]+\Lambda_{ext}.\nonumber \end{align}

Resolviendo este sistema de ecuaciones diferenciales encontramos la
métrica para el exterior del agujero de gusano (ver Apéndice C para
los detalles),\begin{equation}
ds^{2}=-\left(-M-\Lambda_{ext}r^{2}\right)dt^{2}+\frac{dr^{2}}{\left(-M-\Lambda_{ext}r^{2}\right)}+r^{2}d\varphi^{2}.\label{metricaexterior}\end{equation}

La constante cosmológica puede tomar valores positivos o negativos.
En nuestro caso nos restringiremos al caso con $\Lambda_{ext}<0$,
y ya que la constante cosmológica tiene unidades de inverso de longitud
al cuadrado, tomaremos, siguiendo el trabajo de Banados et. al.\cite{BTZ1,BTZ3}

\begin{equation}
\Lambda_{ext}=-\frac{1}{l^{2}}.\end{equation}

Por ello, la solución externa que utilizaremos será

\begin{equation}
ds^{2}=-\left(-M+\frac{r^{2}}{l^{2}}\right)dt^{2}+\frac{dr^{2}}{\left(-M+\frac{r^{2}}{l^{2}}\right)}+r^{2}d\varphi^{2}.\end{equation}

Es importante notar que esta solución tiene singularidades en los
radios

\begin{equation}
r_{\pm}=\pm\sqrt{M}l\end{equation}

El valor $r_{+}$ corresponde al horizonte de eventos de la solución
tipo agujero negro, por lo que para satisfacer las condiciones de
traversabilidad el radio de la boca de nuestro agujero de gusano debe
ser mayor ($a>r_{+}$).

\bigskip

\subsection{Condiciones de Unión}

Para lograr unir las soluciones exterior e interior encontradas debemos
aplicar las condiciones matemáticas necesarias. Si llamamos $S$ la
superficie de frontra entre las regiones exterior e interior, debemos
imponer inicialmente que la métrica sea continua en $S$, es decir
que $\left.g_{\mu v}^{int}\right|_{S}=\left.g_{\mu v}^{ext}\right|_{S}$.
Sin embargo, esta condici\'{o}n no es suficiente para lograr la unión
de las dos regiones. Además de esta, uno de los formalismos desarrollados
impone además la condición de continuidad en la curvatura extrínseca
sobre la superficie $S$ (i.e. la continuidad de la segunda forma
fundamental de $S$). Este es conocido como formalismo de Darmois-Israel.
Afortunadamente, cuando el espacio-tiempo tiene alta simetría (como
en este caso, ya que se posee simetría esférica), la unión puede realizarse
utilizando directamente las ecuaciones de campo. \\

De esta manera encontraremos la densidad de energía y esfuerzos en
la superficie $S$ necesarios para lograr la unión entre las regiones
exterior e interior. Cuando no existen términos de energía ni de esfuerzos
en $S$ se dice que la unión es una \textit{superficie frontera},
mientras que cuando existen términos de energía o esfuerzo, la unión
es llamada un \textit{cascarón-delgado} (\textit{thin-shell}).

\subsubsection{Continuidad de la Métrica}

Ya que tanto la métrica interior como la exterior tienen simetría
esférica, la condición de continuidad de la métrica $\left.g_{\mu v}^{int}\right|_{S}=\left.g_{\mu v}^{ext}\right|_{S}$
está asegurada para la componente $g_{\varphi\varphi}$. De esta manera,
solamente es necesario imponer las condiciones de continuidad\begin{align}
\left.g_{tt}^{int}\right|_{r=a} & =\left.g_{tt}^{ext}\right|_{r=a}\\
\left.g_{rr}^{int}\right|_{r=a} & =\left.g_{rr}^{ext}\right|_{r=a}.\nonumber \end{align}

Utilizando las ecuaciones (\ref{metricainterior}) y (\ref{metricaexterior})
tenemos que estas condiciones de continuidad son\begin{equation}
e^{2\Phi\left(a\right)}=-M+\frac{a^{2}}{l^{2}}\label{aux11}\end{equation}
\begin{equation}
1-\frac{b\left(a\right)}{a}=-M+\frac{a^{2}}{l^{2}},\label{aux8}\end{equation}

las cuales se pueden reescribir como\begin{equation}
\Phi\left(a\right)=\frac{1}{2}\ln\left(-M+\frac{a^{2}}{l^{2}}\right)\end{equation}
\begin{equation}
b\left(a\right)=\left(1+M\right)a-\frac{a^{3}}{l^{2}}.\label{aux12}\end{equation}

De esta última ecuación se puede obtener inmediatamente que la masa
del agujero de gusano estará dada por\begin{equation}
M=\frac{b\left(a\right)}{a}+\frac{a^{2}}{l^{2}}-1.\label{masaagujero}\end{equation}

\subsection{Ecuaciones de Campo}

Para completar la unión de las regiones interior y exterior utilizaremos
las ecuaciones de campo (\ref{eccampo}). Además asumiremos, para
simplificar los cálculos, que los observadores estáticos en el interior
miden fuerzas de marea nulas, es decir que $\Phi^{int}=$constante,
y por lo tanto $\Phi^{\prime int}=0$. Puesto que hemos impuesto la
continuidad de la métrica, y ya que el espacio tiempo es esféricamente
simétrico, solo falta imponer la condición de continuidad en las derivadas
radiales de la métrica. Como sabemos, las segundas derivadas de la
métrica est\'{a}n relacionadas con el tensor de Einstein $G_{\mu v}$,
o en la base ortonormal con $G_{\widehat{\mu}\widehat{v}}$; y además
este es proporcional al tensor momento energía $T_{\widehat{\mu}\widehat{v}}$
por las ecuaciones de campo.

Si las componentes del tensor momento-energía no son nulas en la superficie
$S$ (thin-shell), podremos escribir este tensor de manera proporcional
a un delta de Dirac,\begin{equation}
T_{\widehat{\mu}\widehat{v}}=t_{\widehat{\mu}\widehat{v}}\delta\left(\widehat{r}-\widehat{a}\right),\end{equation}

donde $\widehat{r}=\sqrt{g_{rr}}r$ es la distancia propia medida
dentro del thin-shell. Así, para encontrar $t_{\widehat{\mu}\widehat{v}}$
debemos utilizar\begin{equation}
\int_{int}^{ext}G_{\widehat{\mu}\widehat{v}}d\widehat{r}=\int_{int}^{ext}t_{\widehat{\mu}\widehat{v}}\delta\left(\widehat{r}-\widehat{a}\right)d\widehat{r},\end{equation}

donde $\int_{int}^{ext}$ es una integral infinitesimal a lo largo
de la thin-shell. Utilizando la propiedad de la función delta de Dirac\begin{equation}
\int g\left(x\right)\delta\left(x-x_{o}\right)dx=g\left(x_{o}\right),\end{equation}

tenemos\begin{equation}
t_{\widehat{\mu}\widehat{v}}=\int_{int}^{ext}G_{\widehat{\mu}\widehat{v}}d\widehat{r}.\end{equation}

\subsubsection{Presión Superficial}

Consideraremos inicialmente los términos de densidad de energía superficial
y presión superficial tangencial.

De (\ref{teinstein1}) vemos que la componente $G_{\widehat{t}\widehat{t}}$
depende únicamente de primeras derivadas de la métrica. De esta manera
la densidad de energía superficial será\begin{equation}
\Sigma=t_{\widehat{t}\widehat{t}}=\int_{int}^{ext}G_{\widehat{t}\widehat{t}}d\widehat{r}.\end{equation}

Al realizar la integración se obtendrán solamente funciones de la
métrica que por la condición impuesta serán continuas. Ya que al realizar
la integración se evaluan los términos en la parte exterior e interior,
el resultado sera cero. De esta manera \begin{equation}
\Sigma=0.\end{equation}

\bigskip

Por lo tanto, solamente hace falta encontrar la presión superficial
tangencial $\mathcal{P}$. De las ecuaciones (\ref{teinstein1}) vemos
que la componente $G_{\widehat{\varphi}\widehat{\varphi}}$ tiene
términos que dependen de primeras derivadas de la métrica, y por lo
tanto no contribuyen a la integral; pero tiene además un término de
la forma $\left(1-\frac{b}{r}\right)\Phi^{\prime\prime}$. As\'{\i},
la presión tangencial en la superficie será\begin{equation}
\mathcal{P}=t_{\widehat{\varphi}\widehat{\varphi}}=\int_{int}^{ext}G_{\widehat{\varphi}\widehat{\varphi}}d\widehat{r}\end{equation}
\begin{equation}
\mathcal{P}=\left[\sqrt{1-\frac{b\left(a\right)}{a}}\left.\Phi^{\prime}\right|_{int}^{ext}\right].\end{equation}

Ya que asumimos que un observador estático interno no siente fuerzas
de marea tenemos $\Phi^{\prime int}=0$, y por otro lado tenemos\begin{align}
\Phi^{\prime ext} & =\frac{d}{dr}\left[\frac{1}{2}\ln\left(-M+\frac{r^{2}}{l^{2}}\right)\right]_{r=a}\\
\Phi^{\prime ext} & =\frac{a}{l^{2}}\left(-M+\frac{a^{2}}{l^{2}}\right)^{-1}.\nonumber \end{align}

Utilizando (\ref{aux8}) tenemos\begin{equation}
\Phi^{\prime ext}=\frac{\frac{a}{l^{2}}}{\left(1-\frac{b\left(a\right)}{a}\right)},\label{aux9}\end{equation}

y de esta manera, la presión tangencial en la superficie es

\bigskip\begin{equation}
\mathcal{P}=\frac{\frac{a}{l^{2}}}{\sqrt{1-\frac{b\left(a\right)}{a}}}\label{aux10}\end{equation}

\begin{equation}
\mathcal{P}=\frac{\frac{a}{l^{2}}}{\sqrt{-M+\frac{a^{2}}{l^{2}}}}.\end{equation}

Notese que en este caso, la Presión tangencial en la superficie es
siempre positiva, bajo la condición

\begin{equation}
a^{2}\geq Ml^{2},\end{equation}
la cual corresponde a de cir que la boca del agujero de gusano debe
estar fuera del horizonte de eventos del agujero negro BTZ, tal como
se quiere al buscar un agujero de gusano atravesable.\bigskip

\subsubsection{Presión Radial}

La componente radial de las ecuacones de campo de Einstein (\ref{aux2})
permite escribir para la región interior y exterior\begin{equation}
\tau^{int}\left(r\right)=-\left(1-\frac{b^{int}}{r}\right)\frac{\Phi^{\prime int}}{r}-\Lambda^{int}\end{equation}
\begin{equation}
\tau^{ext}\left(r\right)=-\left(1-\frac{b^{ext}}{r}\right)\frac{\Phi^{\prime ext}}{r}-\Lambda^{ext}\end{equation}

\bigskip

Debido a que hemos asumido que los observadores estáticos en el interior
miden fuerzas de marea nulas, es decir que $\Phi^{\prime int}\left(a\right)=0$,
obtenemos

\begin{equation}
\tau^{int}\left(r\right)=-\Lambda^{int}\end{equation}
\begin{equation}
\tau^{ext}\left(r\right)=-\left(1-\frac{b^{ext}}{r}\right)\frac{\Phi^{\prime ext}}{r}-\Lambda^{ext}\end{equation}

\bigskip

Utilizando la ecuación (\ref{aux9}) para $\Phi^{\prime ext}$ y la
Presión tangencial en la superficie dada por (\ref{aux10}) tenemos

\begin{equation}
\tau^{ext}\left(a\right)=-\left(1-\frac{b^{ext}\left(a\right)}{a}\right)\frac{\frac{a}{l^{2}}}{\left(1-\frac{b^{ext}\left(a\right)}{a}\right)a}-\Lambda^{ext}\end{equation}
\begin{equation}
\tau^{ext}\left(a\right)=-\mathcal{P}\frac{1}{a}\sqrt{1-\frac{b^{ext}\left(a\right)}{a}}-\Lambda^{ext}\end{equation}
\begin{equation}
\tau^{ext}\left(a\right)=-\frac{\mathcal{P}}{a}\sqrt{-M+\frac{a^{2}}{l^{2}}}-\Lambda^{ext}\label{presion radial}\end{equation}

\bigskip

Esta ecuación nos relaciona la tensión radial en la superficie con
la presión tangencial de la thin-shell.

\bigskip

\section{Soluciones Específicas}

Es posible definir varias soluciones que representan agujeros de gusano.
En general estas soluciones puden tener thin-shells en la superficie
donde realizamos la unión entre el exterior y el interior (es decir
que tienen una presion superficial tangencial $\mathcal{P}\neq0$
). Además, asumiremos que en el interior se tiene $\Phi'^{int}=0$
para asegurar la traversabilidad del agujero.

\subsection{Unión con $\mathcal{P}=0$ (Superficie Frontera)}

La soluci\'{o}n exterior tiene $\tau_{ext}=0$ y $\Lambda_{ext}=-\frac{1}{l^{2}}<0$.
Consideraremos primero el caso con $\mathcal{P}=0$. De esta manera,
la ecuación (\ref{presion radial}) será\begin{equation}
\Lambda^{ext}=0.\end{equation}

Esto muestra que no existe solución tipo agujero de gusano con $\mathcal{P}=0$
en universos con constante cosmológica negativa.

\subsection{Unión con $\mathcal{P}\neq0$ (Thin-Shell)}

Tenemos ahora el caso en el que $\tau_{ext}=0$ y $\Lambda_{ext}<0$
pero ahora con una thin-shell, i.e. $\mathcal{P}\neq0$. La ecuaci\'{o}n
(\ref{presion radial}) será ahora\begin{equation}
\frac{\mathcal{P}}{a}\sqrt{-M+\frac{a^{2}}{l^{2}}}=-\Lambda^{ext}=\frac{1}{l^{2}}\end{equation}

y la función de forma estará dada por (\ref{aux12}), es decir\begin{equation}
b\left(a\right)=\left(1+M\right)a-\frac{a^{3}}{l^{2}}\label{aux19}\end{equation}

y por ello la masa del agujero es (\ref{masaagujero})\begin{equation}
M=\frac{b\left(a\right)}{a}+\frac{a^{2}}{l^{2}}-1.\end{equation}

Se observa inmediatamante que la masa es cero si $b\left(a\right)=a-\frac{a^{3}}{l^{2}}$,
es positiva si $b\left(a\right)>a-\frac{a^{3}}{l^{2}}$ y es negativa
cuando $b\left(a\right)<a-\frac{a^{3}}{l^{2}}$.

Para el desarrollo subsecuente, consideraremos el caso límite $b\left(a\right)=a-\frac{a^{3}}{l^{2}}$,
es decir cuando el agujero de gusano no tiene masa. Escogiendo la
función de forma tenemos diferentes agujeros de gusano. Consideraremos
aqui dos posibles funciones.

\bigskip

\begin{enumerate}
\item Consideremos las funciones \begin{align}
b\left(r\right) & =\left(r_{m}r\right)^{\frac{1}{2}}\\
\Phi\left(r\right) & =\Phi_{o}\nonumber \end{align}

donde $r_{m}$ es el radio de la garganta. Con esto tenemos\begin{align}
b^{\prime}\left(r\right) & =\frac{1}{2}\sqrt{\frac{r_{m}}{r}}\\
\Phi^{\prime}\left(r\right) & =0.\nonumber \end{align}

Las ecuaciones de Einstein (\ref{aux14}) a (\ref{aux15}) se podrán
escribir como\begin{equation}
\overline{\rho}\left(r\right)\equiv\rho\left(r\right)+\Lambda_{int}=-\frac{1}{4r^{3}}\sqrt{r_{m}r}\end{equation}
\begin{equation}
\overline{\tau}\left(r\right)\equiv\tau\left(r\right)+\Lambda_{int}=0\end{equation}
\begin{equation}
\overline{p}\left(r\right)\equiv p\left(r\right)-\Lambda_{int}=0.\end{equation}

Por la ecuación (\ref{aux19}) tenemos\begin{equation}
b\left(a\right)=\left(r_{m}a\right)^{\frac{1}{2}}=a-\frac{a^{3}}{l^{2}}\end{equation}
\begin{equation}
\frac{a^{2}}{l^{2}}=1-\left(\frac{r_{m}}{a}\right)^{\frac{1}{2}}.\end{equation}

Para que la solución corresponda a un agujero de gusano y no a un
agujero negro, debemos imponer la condición $a>r_{+}$, con lo que
obtenemos la condición\begin{equation}
a>\sqrt{M}l\end{equation}
\begin{align}
\frac{a^{2}}{l^{2}} & >M\\
1-\left(\frac{r_{m}}{a}\right)^{\frac{1}{2}} & >M\\
\left(\frac{r_{m}}{a}\right)^{\frac{1}{2}} & <1-M\end{align}
\begin{equation}
a>\frac{r_{m}}{\left(M-1\right)^{2}}.\end{equation}

Además, la constante $\Phi_{o}$ debe ser tal que $e^{2\Phi\left(a\right)}=-M+\frac{a^{2}}{l^{2}}$
$\,\,$(ecuación \ref{aux11}), es decir que\begin{equation}
e^{2\Phi_{o}}=-M+\frac{a^{2}}{l^{2}}\end{equation}

Así, la métrica que describe el agujero será: en el interior, ($r_{m}\leq r\leq a$),\begin{equation}
ds^{2}=-\left(-M+\frac{a^{2}}{l^{2}}\right)c^{2}dt^{2}+\frac{dr^{2}}{\left(1-\sqrt{\frac{r_{m}}{r}}\right)}+r^{2}d\varphi^{2}\end{equation}

y en el exterior, ($a\leq r\leq\infty$),\begin{equation}
ds^{2}=-\left(-M+\frac{r^{2}}{l^{2}}\right)dt^{2}+\frac{dr^{2}}{\left(-M+\frac{r^{2}}{l^{2}}\right)}+r^{2}d\varphi^{2}.\end{equation}

\item Otra opción es tomar las funciones \begin{align}
b\left(r\right) & =\frac{r_{m}^{2}}{r}\\
\Phi\left(r\right) & =\Phi_{o}\nonumber \end{align}

con $r_{m}$ el radio de la garganta. Tenemos ahora\begin{align}
b^{\prime}\left(r\right) & =-\frac{r_{m}^{2}}{r^{2}}\\
\Phi^{\prime}\left(r\right) & =0.\nonumber \end{align}

Las ecuaciones de campo están dadas por\begin{equation}
\overline{\rho}\left(r\right)\equiv\rho\left(r\right)+\Lambda_{int}=\frac{1}{2r^{3}}\left[-\frac{r_{m}^{2}}{r^{2}}r-\frac{r_{m}^{2}}{r}\right]=-\frac{r_{m}^{2}}{r^{5}}\end{equation}
\begin{equation}
\overline{\tau}\left(r\right)\equiv\tau\left(r\right)+\Lambda_{int}=0\end{equation}
\begin{equation}
\overline{p}\left(r\right)\equiv p\left(r\right)-\Lambda_{int}=0.\end{equation}

Ahora bien, debido a la ecuación (\ref{aux16}) tenemos\begin{equation}
b\left(a\right)=\frac{r_{m}^{2}}{a}=a-\frac{a^{3}}{l^{2}}\end{equation}
\begin{equation}
r_{m}^{2}=a^{2}-\frac{a^{4}}{l^{2}}.\end{equation}

Es decir que la boca del agujero de gusano debe estar en\begin{equation}
a^{2}=\frac{l^{2}}{2}\left[1\pm\sqrt{1-4\frac{r_{m}^{2}}{l^{2}}}\right].\end{equation}

Para que la solución corresponda a un agujero de gusano debemos imponer
la condición $a>r_{+}$, de donde tenemos\[
a^{2}>Ml^{2}\]
\begin{equation}
\frac{l^{2}}{2}\left[1\pm\sqrt{1-4\frac{r_{m}^{2}}{l^{2}}}\right]>Ml^{2}\end{equation}
\begin{equation}
1\pm\sqrt{1-4\frac{r_{m}^{2}}{l^{2}}}>2M.\end{equation}

La constante $\Phi_{o}$ debe ser tal que $e^{2\Phi\left(a\right)}=-M+\frac{a^{2}}{l^{2}}$
$\,\,$(ecuación \ref{aux11}), por locual\begin{equation}
e^{2\Phi_{o}}=-M+\frac{a^{2}}{l^{2}}.\end{equation}

Así, la métrica en el interior del agujero de gusano, $r_{m}\leq r\leq a$,
está dada por\begin{equation}
ds^{2}=-\left(-M+\frac{a^{2}}{l^{2}}\right)c^{2}dt^{2}+\frac{dr^{2}}{\left(1-\frac{r_{m}^{2}}{r^{2}}\right)}+r^{2}d\varphi^{2}\end{equation}

y la m\'{e}trica exterior, $a\leq r\leq\infty,$ será\begin{equation}
ds^{2}=-\left(-M+\frac{r^{2}}{l^{2}}\right)dt^{2}+\frac{dr^{2}}{\left(-M+\frac{r^{2}}{l^{2}}\right)}+r^{2}d\varphi^{2}.\end{equation}

\end{enumerate}
\bigskip

\section{Conclusiones}

En este trabajo hemos considerado el método usual de construcción
de agujeros de gusano, realizando la unión de dos espacio-tiempo dentro
del formalismo de la gravedad (2+1) en un universo con constante cosmológica
negativa. En la región interna se exige una métrica que posea las
características geométricas y de traversabilidad deseadas para el
agujero de gusano, mientras que en la región exterior se exige que
la métrica corresponda al agujero negro BTZ. Con ello se logran encontrar
explícitamente dos soluciones específicas que corresponden a agujeros
de gusano atravesables y que en su región exterior corresponden a
soluciones BTZ. Por último, del análisis realizado, se muestra que
las dos soluciones específicas encontradas necesitan de materia exótica
para mantenerse.

\appendix

\section*{Apéndice A. Tensores para la métrica de Agujero de Gusano}

Para el cálculo de los tensores de Riemann y de Einstein a partir
de la métrica se utilizó el paquete grTensor para Mathematica 5.0.

Para nuestra métrica (\ref{metrica}), escrita como:\begin{equation}
ds^{2}=g_{\alpha\beta}dx^{\alpha}dx^{\beta}=-e^{2\Phi\left(r\right)}dt^{2}+\frac{1}{1-\frac{b\left(r\right)}{r}}dr^{2}+r^{2}d\varphi^{2}\end{equation}

los simbolos de Christoffel vienen dados por:\begin{equation}
\Gamma_{\beta\gamma}^{\alpha}=\frac{1}{2}g^{\alpha\lambda}\left(g_{\lambda\beta,\gamma}+g_{\lambda\gamma,\beta}-g_{\gamma\beta,\lambda}\right)\end{equation}

donde hemos utilizado la coma para denotar las derivadas parciales,
i.e. $g_{\alpha\beta,\gamma}=\frac{\partial g_{\alpha\beta}}{\partial x^{\gamma}}.$

De esta forma, las componentes no nulas de los simbolos de Christoffel
son\begin{align*}
\Gamma_{tt}^{r} & =\left(1-\frac{b}{r}\right)\Phi^{\prime}e^{2\Phi}\\
\Gamma_{tr}^{t} & =\Phi^{\prime}\\
\Gamma_{rr}^{r} & =\frac{1}{2}\frac{\left(b^{\prime}-\frac{b}{r}\right)}{r-b}\\
\Gamma_{r\varphi}^{\varphi} & =\frac{1}{r}\\
\Gamma_{\varphi\varphi}^{r} & =b-r\end{align*}

\bigskip

Donde las primas denotan derivadas con respecto a la coordenada $r$.
El tensor de Riemann se calcula a partir de los simbolos de Christoffel
mediante:\begin{equation}
R_{\beta\gamma\delta}^{\alpha}=\Gamma_{\beta\delta,\gamma}^{\alpha}-\Gamma_{\beta\gamma,\delta}^{\alpha}+\Gamma_{\lambda\gamma}^{\alpha}\Gamma_{\beta\delta}^{\lambda}-\Gamma_{\lambda\delta}^{\alpha}\Gamma_{\beta\gamma}^{\lambda}\end{equation}

Obtenemos entonces componentes no nulas:\begin{align*}
R_{rtr}^{t} & =\frac{\Phi^{\prime}}{2r}\left(\frac{b-b^{\prime}r}{b-r}\right)-\left(\Phi^{\prime}\right)^{2}-\Phi^{\prime\prime}\\
R_{\varphi t\varphi}^{t} & =\left(b-r\right)\Phi^{\prime}\\
R_{ttr}^{r} & =-\frac{e^{2\Phi}}{2r^{2}}\left[\Phi^{\prime}\left(b-b^{\prime}r\right)+2r\left(r-b\right)\left(\left(\Phi^{\prime}\right)^{2}+\Phi^{\prime\prime}\right)\right]\\
R_{\varphi r\varphi}^{r} & =\frac{\left(b^{\prime}r-b\right)}{2r}\\
R_{tt\varphi}^{\varphi} & =\frac{e^{2\Phi}\left(b-r\right)\Phi^{\prime}}{r^{2}}\\
R_{rr\varphi}^{\varphi} & =\frac{\left(b-b^{\prime}r\right)}{2r^{2}\left(r-b\right)}\end{align*}

\bigskip

A partir del tensor de Riemann construimos el tensor de Ricci, dado
por:\begin{equation}
R_{\alpha\beta}=R_{\alpha\gamma\beta}^{\gamma}\end{equation}

Para nuestra métrica tenemos las componentes:\begin{align*}
R_{tt} & =\frac{e^{2\Phi}}{2r^{2}}\left[\Phi^{\prime}\left(2r-b-b^{\prime}r\right)+2r\left(r-b\right)\left(\left(\Phi^{\prime}\right)^{2}+\Phi^{\prime\prime}\right)\right]\\
R_{rr} & =\frac{1}{2r^{2}\left(b-r\right)}\left[\left(b-b^{\prime}r\right)\left(1+r\Phi^{\prime}\right)+2r^{2}\left(r-b\right)\left(\left(\Phi^{\prime}\right)^{2}+\Phi^{\prime\prime}\right)\right]\\
R_{\varphi\varphi} & =\frac{1}{2r}\left[2\Phi^{\prime}r\left(b-r\right)+b^{\prime}r-b\right]\end{align*}

\bigskip

La contracción del Tensor de Ricci produce el Escalar de Curvatura,\begin{equation}
R=R_{\alpha}^{\alpha}.\end{equation}

En este caso, este tiene el valor\[
R=\frac{1}{r^{3}}\left[2\Phi^{\prime\prime}r^{2}\left(b-r\right)+\Phi^{\prime}r\left(b^{\prime}r-2r+b\right)+2\Phi^{\prime2}r^{2}\left(b-r\right)+b^{\prime}r-b\right].\]

\bigskip

Por \'{u}ltimo, utilizando los tensores anteriores, construimos el
tesnor de Eintein, definido por

\bigskip\begin{equation}
G_{\alpha\beta}=R_{\alpha\beta}-\frac{1}{2}g_{\alpha\beta}R.\end{equation}

Las componentes no nulas del tensor de Einstein son entonces\begin{align*}
G_{tt} & =\frac{1}{2r^{3}}e^{2\Phi}\left[-b+rb^{\prime}\right]\\
G_{rr} & =\frac{\Phi^{\prime}}{r}\\
G_{\varphi\varphi} & =\frac{1}{2}\left[\Phi^{\prime}\left(b-rb^{\prime}\right)+2r\left(r-b\right)\left(\left(\Phi^{\prime}\right)^{2}+\Phi^{\prime\prime}\right)\right].\end{align*}

\bigskip

\section*{Apéndice B. Cambio de Base para el Tensor de Einstein}

\bigskip

La base asociada con las coordenadas $\left(t,r,\varphi\right)$ es
denotada por $\left(\mathbf{e}_{t},\mathbf{e}_{r},\mathbf{e}_{\varphi}\right)$
o en el lenguaje de la geometría diferencial como:\begin{align*}
\mathbf{e}_{t} & =\frac{\partial}{\partial t}\\
\mathbf{e}_{r} & =\frac{\partial}{\partial r}\\
\mathbf{e}_{\varphi} & =\frac{\partial}{\partial\varphi}\end{align*}

La base de vectores ortonormales que utilizaremos est\'{a} definida
por la transformación : $\mathbf{e}_{\widehat{\alpha}}=\Lambda_{\widehat{\alpha}}^{\beta}\mathbf{e}_{\beta}$
, donde\begin{equation}
\Lambda_{\widehat{\alpha}}^{\beta}=diag\left[e^{-\Phi},\left(1-\frac{b}{r}\right)^{1/2},\frac{1}{r}\right]\end{equation}

Explícitamente tenemos:\begin{align*}
\mathbf{e}_{\widehat{t}} & =e^{-\Phi}\mathbf{e}_{t}\\
\mathbf{e}_{\widehat{r}} & =\left(1-\frac{b}{r}\right)^{1/2}\mathbf{e}_{r}\\
\mathbf{e}_{\widehat{\varphi}} & =\frac{1}{r}\mathbf{e}_{\varphi}\end{align*}

De esta manera para cambiar el tensor de Einstein calculado en el
Apéndice A a la base ortonormal debemos utilizar la transformación:\begin{equation}
G_{\widehat{\alpha}\widehat{\beta}}=\Lambda_{\widehat{\alpha}}^{\mu}\Lambda_{\widehat{\beta}}^{v}G_{\mu v}\end{equation}

Tenemos entonces explícitamente:\begin{align*}
G_{\widehat{t}\widehat{t}} & =\Lambda_{\widehat{t}}^{\mu}\Lambda_{\widehat{t}}^{v}G_{\mu v}=\Lambda_{\widehat{t}}^{t}\Lambda_{\widehat{t}}^{t}G_{tt}\\
 & =e^{-\Phi}e^{-\Phi}\frac{1}{2r^{3}}e^{2\Phi}\left[-b+rb^{\prime}\right]\\
 & =\frac{1}{2r^{3}}\left[b^{\prime}r-b\right]\\
 & \text{ \,\,\,\,}\\
G_{\widehat{r}\widehat{r}} & =\Lambda_{\widehat{r}}^{\mu}\Lambda_{\widehat{r}}^{v}G_{\mu v}=\Lambda_{\widehat{r}}^{r}\Lambda_{\widehat{r}}^{r}G_{rr}\\
 & =\left(1-\frac{b}{r}\right)\frac{\Phi^{\prime}}{r}\\
 & \text{ \,\,\,\,}\\
G_{\widehat{\varphi}\widehat{\varphi}} & =\Lambda_{\widehat{\varphi}}^{\mu}\Lambda_{\widehat{\varphi}}^{v}G_{\mu v}=\Lambda_{\widehat{\varphi}}^{\varphi}\Lambda_{\widehat{\varphi}}^{\varphi}G_{\varphi\varphi}\\
 & =\left(\frac{1}{r^{2}}\right)\frac{1}{2}\left[\Phi^{\prime}\left(b-rb^{\prime}\right)+2r\left(r-b\right)\left(\left(\Phi^{\prime}\right)^{2}+\Phi^{\prime\prime}\right)\right]\\
 & =\frac{1}{2r^{2}}\left[\Phi^{\prime}\left(b-rb^{\prime}\right)+2r\left(r-b\right)\left(\left(\Phi^{\prime}\right)^{2}+\Phi^{\prime\prime}\right)\right]\end{align*}

\bigskip

De esta manera, al cambiar de base, las nuevas componentes del tensor
de Einstein son finalmente:\begin{align*}
G_{\widehat{t}\widehat{t}} & =\frac{1}{2r^{3}}\left[b^{\prime}r-b\right]\\
G_{\widehat{r}\widehat{r}} & =\left(1-\frac{b}{r}\right)\frac{\Phi^{\prime}}{r}\\
G_{\widehat{\varphi}\widehat{\varphi}} & =\frac{1}{2r^{2}}\left[\Phi^{\prime}\left(b-rb^{\prime}\right)+2r\left(r-b\right)\left(\left(\Phi^{\prime}\right)^{2}+\Phi^{\prime\prime}\right)\right]\end{align*}

\section*{Apéndice C. Métrica para el Exterior del Agujero de Gusano}

\bigskip

Para el exterior del agujero de gusano consideramos un espacio-tiempo
vacio, por lo que utilizamos un tensor momento-energía nulo $T_{\widehat{\mu}\widehat{v}}=0$.
Esta condición se reduce a tomar:\[
\rho\left(r\right)=\tau\left(r\right)=p\left(r\right)=0\]

De esta manera las ecuaciones de campo se convierten en:\begin{align*}
0 & =\frac{1}{2r^{3}}\left[b^{\prime}r-b\right]-\Lambda_{ext}\\
0 & =-\left(1-\frac{b}{r}\right)\frac{\Phi^{\prime}}{r}-\Lambda_{ext}\\
0 & =\frac{1}{2r^{2}}\left[\Phi^{\prime}\left(b-rb^{\prime}\right)+2r\left(r-b\right)\left(\left(\Phi^{\prime}\right)^{2}+\Phi^{\prime\prime}\right)\right]+\Lambda_{ext}\end{align*}

De la primera de estas ecuaciones tenemos la ecuacion diferencial\[
\frac{1}{2r^{3}}\left[b^{\prime}r-b\right]=\Lambda_{ext},\]

de la cual, obtenemos por integración\begin{equation}
b\left(r\right)=\Lambda_{ext}r^{3}+Kr\end{equation}

donde $K$ es una constante. Reemplazando en la segunda ecuación tenemos:\begin{align*}
0 & =-\left(1-\frac{\Lambda_{ext}r^{3}+Kr}{r}\right)\frac{\Phi^{\prime}}{r}-\Lambda_{ext}\\
0 & =-\left(1-\Lambda_{ext}r^{2}+K\right)\frac{\Phi^{\prime}}{r}-\Lambda_{ext}\end{align*}

Es decir:\[
\Phi^{\prime}=-\frac{r\Lambda_{ext}}{\left(1-\Lambda_{ext}r^{2}+K\right)}\]
\[
\Phi=-\int\frac{r\Lambda_{ext}}{\left(1-\Lambda_{ext}r^{2}+K\right)}dr\]

Haciendo el cambio de variable: $u=1-\Lambda_{ext}r^{2}+K$ tenemos:\begin{align*}
\Phi & =\frac{1}{2}\int\frac{1}{u}du=\frac{1}{2}\ln u+K_{1}\\
\Phi\left(r\right) & =\frac{1}{2}\ln\left(1-\Lambda_{ext}r^{2}+K\right)+K_{1}\end{align*}

donde $K_{1}$ es otra cosntante. De esta manera, el elemento de linea
en el exterior del agujero tendr\'{a} la forma:\[
ds^{2}=-e^{2\Phi\left(r\right)}dt^{2}+\frac{1}{1-\frac{b\left(r\right)}{r}}dr^{2}+r^{2}d\varphi^{2}\]
\[
ds^{2}=-e^{2\left[\frac{1}{2}\ln\left(1-\Lambda_{ext}r^{2}+K\right)+K_{1}\right]}dt^{2}+\frac{dr^{2}}{\left(1-\frac{\Lambda_{ext}r^{3}+Kr}{r}\right)}+r^{2}d\varphi^{2}\]
\[
ds^{2}=-e^{2K_{1}}\left(1-\Lambda_{ext}r^{2}+K\right)dt^{2}+\frac{dr^{2}}{\left(1-\Lambda_{ext}r^{2}+K\right)}+r^{2}d\varphi^{2}\]

\bigskip

Ahora bien, al comparar con la conocida solución BTZ se observa que
las constantes de integración que aparecen deben ser\[
K=-M-1\text{ \,\,\,\,\,\,\,\,\,\,\,\,\,\,\,\,\,\,\,\,\,\,\,\,\,\,\,\,\,\,\,\,\,\,\,\,}K_{1}=0\]

Por lo tanto la métrica exterior será finalmente\[
ds^{2}=-\left(-M-\Lambda_{ext}r^{2}\right)dt^{2}+\frac{dr^{2}}{\left(-M-\Lambda_{ext}r^{2}\right)}+r^{2}d\varphi^{2}\]

$\allowbreak$

\end{document}